%% file: tyumenev_etal_comparingHg.tex
\documentclass[12pt]{iopart}
\usepackage[dvips]{graphicx}
\graphicspath{./}
\usepackage{subfig}
\usepackage[dvips]{xcolor}
\usepackage{framed, caption}
\usepackage{cleveref}
\usepackage[normalem]{ulem}
\newcommand{\erec}{\ensuremath{E_\mathrm{rec}}}
\newcommand{\numagic}{\ensuremath{\nu_\mathrm{magic}}}

\newcommand{\tm}[1]{\ensuremath{10^{-#1}}}

\usepackage[sorting=none,backend=bibtex,url=false,isbn=false]{biblatex}
\bibliography{bibliography}

\begin{document}
\title[Comparing a mercury OL clock with microwave and optical frequency standards]{Comparing a mercury optical lattice clock with microwave and optical frequency standards}

\author{R Tyumenev, M Favier, S Bilicki, E Bookjans, R Le Targat, J Lodewyck, D Nicolodi, Y Le Coq, M Abgrall, J Gu\'ena, L De Sarlo, S Bize}
\address{LNE-SYRTE, Observatoire de Paris, PSL Research University, CNRS, Sorbonne Universit\'es, UPMC Univ. Paris 06, 61 avenue de l'Observatoire, 75014 Paris, France}

\begin{abstract}
In this paper we report the evaluation of an optical lattice clock based on neutral mercury with a relative uncertainty of $1.7\times\tm{16}$. Comparing this characterized frequency standard to a $^{133}$Cs atomic fountain we determine the absolute frequency of the $^1S_0 \rightarrow \phantom{}^3P_0$ transition of $^{199}$Hg as $\nu_{\mathrm{Hg}} = 1 128\,575\,290\,808\,154.62\,$Hz $\pm\,0.19\,$Hz (statistical) $\pm\,0.38\,$Hz (systematic), limited solely by the realization of the SI second.
Furthermore, by comparing the mercury optical lattice clock to a $^{87}$Rb atomic fountain, we determine for the first time to our knowledge the ratio between the $^{199}$Hg  clock transition and the $^{87}$Rb ground state hyperfine transition.
Finally we present a direct optical to optical measurement of the $^{199}$Hg/$^{87}$Sr frequency ratio. The obtained value of $\nu_{\mathrm{Hg}}/\nu_{\mathrm{Sr}}=$2.629~314~209~898~909~15 with a fractional uncertainty of $1.8\times\tm{16}$ is in excellent agreement with a similar measurement obtained by Yamanaka \emph{et al.} [Phys. Rev. Lett. {\bf 114} 230801 (2015)]. This makes this frequency ratio one of the few physical quantities agreed upon by different laboratories to this level of uncertainty. Frequency ratio measurements of the kind reported in this paper have a strong impact for frequency metrology and fundamental physics as they can be used to monitor putative variations of fundamental constants.
\end{abstract}

\section{Introduction}\label{sec:intro}
\input{intro}
\section{The mercury optical lattice clock setup}\label{sec:molc}
\input{molc}
\section{Uncertainty budget}\label{sec:systematics}
\input{systematics}
\section{Frequency ratio measurements}\label{sec:ratio}
\input{comparison}
\section{Acknowledgment}
\input{acknowledgment}

\printbibliography

\end{document}

%% file: intro.tex
The emergence of optical lattice clocks has sustained a trend of increasing accuracy in the measurement of frequencies at a pace even faster than that observed since the inception of atomic fountains twenty years ago (see \cite{Ludlow2015} and references therein). The possibility of interrogating thousands of neutral atoms suppressing motional effects in a trap that does not perturb the clock transition measurement \cite{Katori2003} enables new applications of atomic clocks such as chronometric geodesy \cite{Chou2010,Lisdat2015} and poses the question of a new definition of the SI unit of time, based on an optical transition, in such concrete terms that a ``Roadmap to the redefinition of the second'' is being drafted by the International Committee for Weights and Measures \cite{CCTF2015}.

A prerequisite for application of optical lattice clocks to both fundamental physics and metrology is confidence in the evaluation of the systematic shifts affecting the measurement of the clock frequency. Such a confidence can only be fostered by repeated comparisons with other frequency standards both in the microwave and in the optical domain \cite{Akamatsu2014,Takamoto2015,Yamanaka2015,Nemitz2016}. Among all possible frequency ratio measurements, a particular importance has to be given to those ratios involving the present definition of the SI second, namely $^{133}$Cs ground state hyperfine splitting, or the transitions recognized as secondary representations of the second by the CIPM \cite{BIPM_LOR}.

In this paper we report the measurement of three of such frequency ratios involving an optical lattice clock based on mercury and a primary frequency standard (Cs fountain), and two standards based on a secondary representation of the second operating respectively in the optical ($\phantom{}^{87}$Sr $\phantom{}^1S_0 \rightarrow \phantom{}^3P_0$ transition) and in the microwave domain ($^{87}$Rb ground state hyperfine transition). In particular we report the first optical to optical reproduction of a frequency ratio between two optical clocks (Hg/Sr) to an accuracy beyond that of the realization of the SI second, in good agreement with a previous measurement \cite{Yamanaka2015}. We also report the first direct measurement of the Hg/Rb frequency ratio with a better accuracy than the one available from indirect measurements.

The interest of these measurements lies beyond the domain of time and frequency metrology as these frequency ratios can be used for fundamental studies. The most prominent example is the search for a time or gravity-related variation of the fine structure constant and of the proton to electron mass ratio (see for instance \cite{Guena2012,Uzan2011,Will2006}).

The structure of the paper is the following: after describing the experimental setup of the mercury optical lattice clock in section \ref{sec:molc}, we present in section \ref{sec:systematics} an evaluation of systematic effects affecting the clock frequency measurement at the level of $1.7\times\tm{16}$. In section \ref{sec:ratio} we then present the frequency comparison infrastructure and the obtained results discussing how they compare to similar measurements known in the literature.

%% file: molc.tex
\begin{figure}[htbp]
	\centering
	\includegraphics[width=0.9\textwidth]{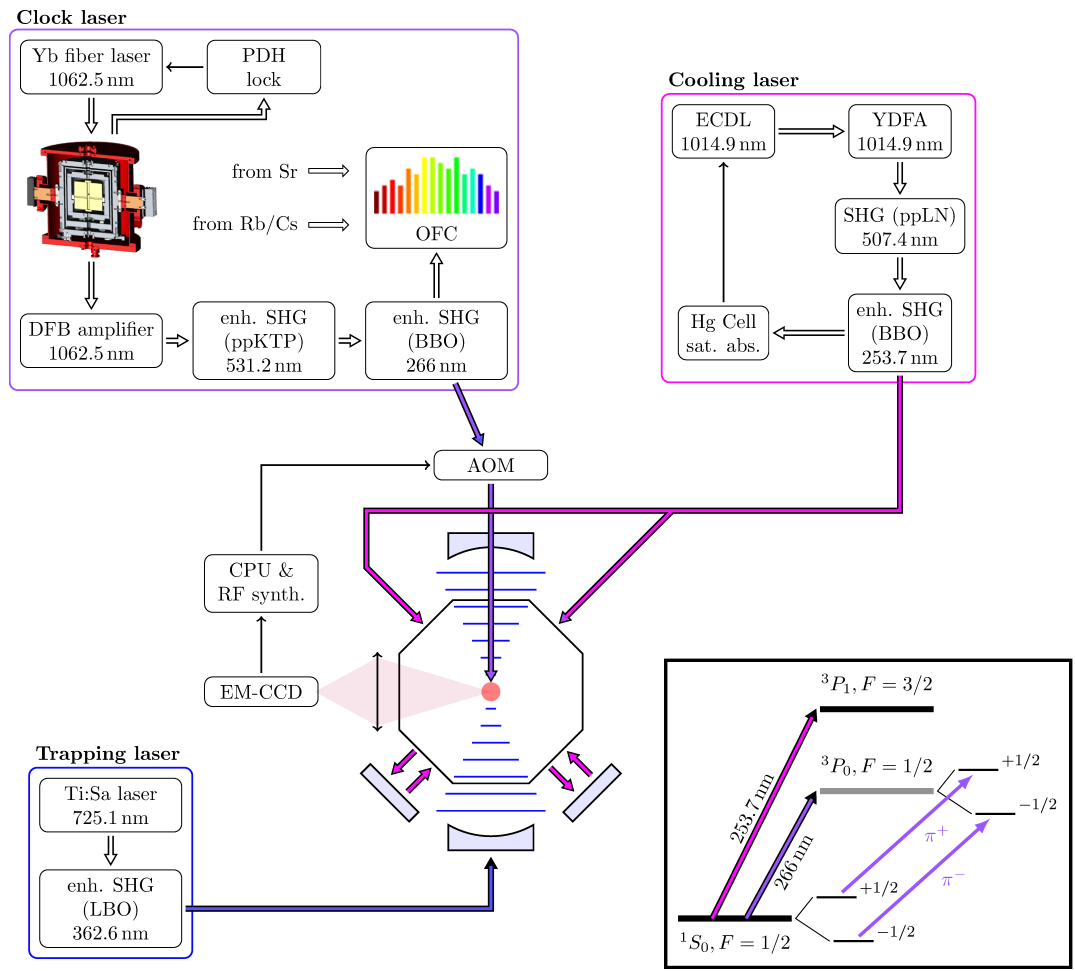}
	\caption{Mercury optical lattice clock experimental setup. Mercury atoms are trapped from a thermal atomic beam and cooled in a MOT (only two pairs of beams shown). The cooling light is generated by a laser system based on an external cavity diode laser source (ECDL) which is frequency stabilized via saturated absorption spectroscopy in a mercury vapor cell and amplified in an Ytterbium-doped fiber amplifier (YDFA). The lattice trap is formed in a Fabry-Perot cavity using light from a home made Ti:Sa laser frequency doubled in a LBO non-linear crystal. The trapped atoms are probed by the clock light pulsed by an acousto-optic modulator (AOM), and subsequently detected via fluorescence at the cooling light frequency by an electron-multiplied low noise CCD camera. Inset shows a level scheme of mercury, note the two $\pi$ components of the clock transition.}
	\label{fig:MOLCfigure1}
\end{figure}

As mentioned in the introduction, the operation of an optical lattice clock requires cooling, trapping and interrogation of an atomic sample. One of the remarkable properties of mercury is that all these operations can be performed with only three laser sources. This is due to a simple level structure depicted in the inset of figure \ref{fig:MOLCfigure1}. A relatively strong ($\Gamma \simeq 2\pi \times 1.2\,$MHz linewidth) intercombination transition $^1S_0\rightarrow \phantom{}^3P_1$ around $254\,$nm allows sub-Doppler laser cooling of fermion isotopes to temperatures of $\sim 30$ $\mu$K in a Magneto-Optical Trap (MOT) \cite{McFerran2010}. This relatively low cooling temperature allows us to load mercury atoms in a dipole trap from a single stage MOT. The magic wavelength for which the trapping light shift is minimized is around $362\,$nm \cite{Yi2011}.
Trapped atoms are then probed by a clock laser tuned on the $^1S_0\rightarrow \phantom{}^3P_0$ doubly forbidden, slightly allowed by hyperfine mixing, intercombination line around $266\,$nm or $1.129\,$PHz. The production of these three uncommon, deep UV laser light with sufficient power and reliability is the major challenge in the design and operation of the mercury optical lattice clock.

A schematic of the clock is shown in figure \ref{fig:MOLCfigure1}. It consists of four subsystems: a vacuum chamber required to provide an ultra-high vacuum environment with a residual pressure of mercury around $10^{-9}\,$mbar and the three laser subsystems mentioned before.

The vacuum apparatus of the experiment consists of two sections. The first section contains the source of mercury atoms, which is a small droplet of mercury kept in a copper tube at -40 $^{\circ}$C. The second section contains the 3D-MOT and a vertically oriented 1D optical lattice trap. The two sections are separated by a 1.5 mm diameter 12 mm long tube, which forms an atomic beam and creates differential pressure of 3 orders of magnitude between the two chambers. A detailed description of the vacuum apparatus can be found in \cite{Petersen2009}.

\subsection{Cooling laser and magneto-optical trapping}
The cooling laser source at $254\,$nm is based on an external cavity diode laser (ECDL), generating light at $1015\,$nm with a linewidth below $10\,$kHz on a millisecond timescale. Approximately $15\,$mW of the output of the ECDL are coupled to an Yb doped fiber amplifier obtaining an output power in excess of $10\,$W. The infrared light generated by the amplifier is subsequently frequency doubled twice, first in a single-pass configuration in a periodically poled Lithium Niobate (ppLN) crystal and then in an angle-tuned $\beta$-barium borate (BBO) crystal in an enhancement configuration, using a bow-tie cavity.

Based on our past experience with BBO crystals, the enhancement cavity and the BBO crystal are placed in a quasi-sealed enclosure under oxygen overpressure. This helps prevent UV-induced damage on the crystal and optical elements' anti-reflection coatings. The enclosure was made out of a single aluminium alloy block. The geometry of the doubling cavity was chosen to have a relatively large waist of 30 $\mu$m, as a compromise between conversion efficiency and long-term damage of the crystal caused by the UV light. This turned out to be instrumental in improving the reliability of our cooling laser source and allowed us to achieve stable and reliable generation of typically $70\,$mW at $254\,$nm. With this amount of cooling light we routinely trap $\sim$10$^6$ atoms in a 3D-MOT with a loading time of $0.7\,$s.

The 3D-MOT is formed by the intersection of three orthogonal pairs of retro-reflected beams red-detuned by $1.5\,\Gamma$ with respect to the cooling transition, and a quadrupole magnetic field with a gradient of $~150\, \mu\mathrm{T}/\mathrm{mm}$ at the center of the trap \cite{Mejri2011}. The MOT is formed in presence of the lattice light as the differential light shift on the cooling transition is typically much smaller than the applied detuning. After the MOT is formed, the cooling light and magnetic field are turned off and atoms are released in free fall. During this phase a low-energy fraction of the MOT atoms gets trapped into the lattice.

\subsection{Lattice trap}\label{sec:lattice}
The lattice trap is formed in a Fabry-Perot (FP) build-up cavity whose waist is centered on the MOT well within one Rayleigh range ($\sim4\,$cm). Mirrors of the Fabry-Perot cavity are mounted with indium UHV seals on the 3D-MOT chamber and are therefore under vacuum.

Mercury has a low polarisability at the magic wavelength and therefore requires high intensity light to create a lattice deep enough to trap atoms well in the Lamb-Dicke regime. At a constant optical power circulating in the build-up cavity of $5.5\,$W, limited by the available laser power at the wavelength of $362.5\,$nm, one needs to find a compromise between increasing the intensity and limiting the long term damage to optics due to exposure to UV light. For this reason, our trap has a mirror radius of curvature of $150\,$mm corresponding to a calculated waist of 69 $\mu$m.
The finesse of the build-up cavity was measured to be about 300. The depth of the lattice trap corresponding to the nominal circulating power was measured to be $\sim 60$\erec, where $\erec\simeq h\times7.57\,$kHz is the kinetic energy of a mercury atom with a momentum equal to that of a lattice photon. In agreement with a simple model of the lattice loading process, this trap allows us to typically trap $\sim 5 \times 10^3$ atoms.

\begin{figure}[htbp]
	\centering
	\includegraphics[width=.6\linewidth]{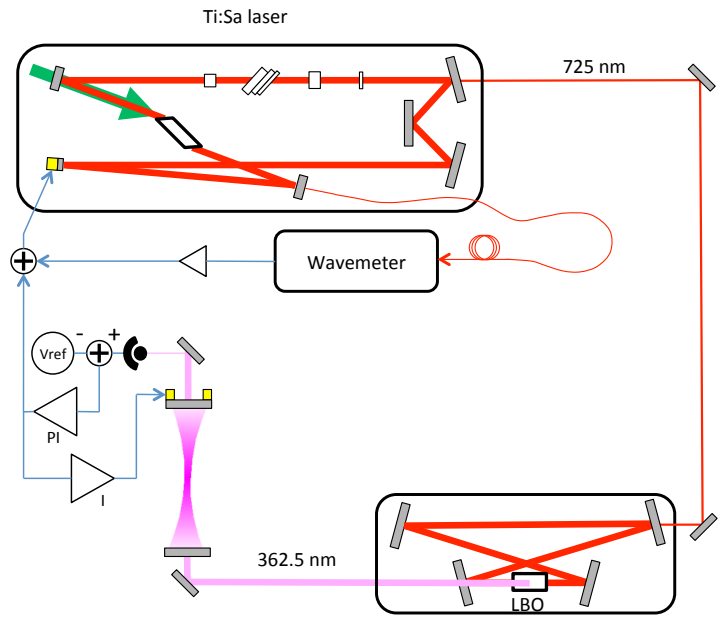}
	\caption{Scheme of the lattice light laser source. The Ti:Sa output light at 725nm is frequency doubled in a LBO non-linear crystal and coupled to the lattice cavity. A system of frequency lock based on a high precision commercial wavemeter and three feedback loops allows us to lock the lattice cavity on the frequency of the Ti:Sa laser and the frequency of the laser to the magic frequency (see text for details).}
	\label{fig2}
\end{figure}

The lattice light is generated by coupling a frequency doubled titanium-sapphire (Ti:Sa) laser to the vertical Fabry-Perot cavity. The laser is a home made non-planar ring cavity laser and frequency doubling is obtained with a bow-tie enhancement cavity in a lithium triborate (LBO) crystal \cite{McFerran2014}. A scheme of the lattice laser system is shown in figure \ref{fig2}.

In order to minimize lattice light shifts, we actively stabilize the lattice wavelength around the ``magic frequency'' (see section \ref{sec:lattice_shift}) through a control scheme working as follows.
The light emitted by the Ti:Sa laser is monitored by a commercial wavemeter with a frequency accuracy of $6\,$MHz and a rate of 1 measurement per second. The wavemeter is calibrated according to the manufacturer specification using the clock light of a Strontium optical lattice clock at $698\,$nm. The error signal derived from this measurement is integrated and fed to the Ti:Sa cavity length to keep its frequency at the desired value with a typical timescale of $5\,$s. A fast PI controller based on an error signal obtained with the H\"ansch-Couillaud method \cite{Haensch1980} keeps the doubling stage locked onto the emission of the Ti:Sa laser.
To obtain stable optical power build-up in the lattice cavity, it would
be necessary to lock the cavity length to the lattice light frequency. However, to overcome the bandwidth limitation imposed by the cavity length piezoelectric actuator, we need to resort to a more complicated scheme, depicted in figure \ref{fig2}, where the Ti:Sa wavelength follows the lattice cavity length on short timescale, and vice-versa for timescales above roughly $10\,$ms. This is achieved in the following way: the relative frequency fluctuations between the Ti:Sa laser and the cavity are measured with a bandwidth of around $10\,$kHz by observing the transmission of the cavity at half-fringe. These fluctuations are processed by a PI controller and fed back to the length of the Ti:Sa cavity, summing this fast control to the slow steering obtained from the wavemeter, as described above. The residual error signal is integrated further and fed to the FP lattice cavity length piezo actuator with a bandwidth of typically $30\,$Hz. The in-loop frequency fluctuations of the locked lattice laser are averaged down to less than 1 MHz at 100s.

\subsection{Clock laser}\label{sec:clock}
The clock laser is based on a commercial Yb fiber laser whose frequency is stabilized on the main mode of a 10 cm long vertically oriented Fabry-Perot cavity with finesse $\sim$850 000 using the Pound-Drever-Hall method \cite{Drever1983}.
The ultra-stable light is frequency shifted by an AOM, amplified in an injection-locked DFB laser and frequency doubled twice in two separate enhancement cavities containing respectively a ppKTP (for 531 nm) and a BBO crystal (for 266 nm). The AOM is also used to stabilize the optical path between the Fabry-Perot cavity and the first enhancement cavity. Spectroscopy and state selection (see Section \ref{sec:state_selection}) pulses are formed by pulsing a second AOM and delivered to the atoms along the lattice axis (see figure \ref{fig:MOLCfigure1}).
The clock laser features a frequency flicker noise floor around $4 \times \tm{16}$ and a typical frequency drift of $20\,$mHz/s \cite{McFerran2012a}. More details can be found in \cite{Millo2009}.

\subsection{State selection, spectroscopy of the clock transition and clock operation}\label{sec:state_selection}
\begin{figure}[htbp]
	\centering
	\includegraphics[width=.8\textwidth]{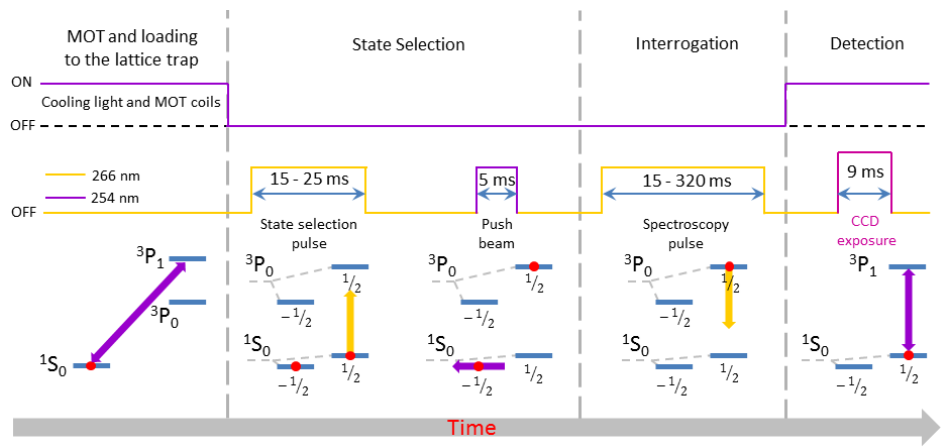}
	\caption{Time sequence of the spectroscopy experiment performed with state selection. Here, the purple and yellow colors indicate the cooling and probe light respectively. After the MOT loading phase, the MOT cooling beams and MOT coils are turned off and a fraction of the MOT atoms are trapped in the lattice. During the state selection phase the state $|^3P_0\rangle$ is populated by the state selection pulse. Subsequently the atoms in the $|^1S_0,m_F=\pm 1/2\rangle$ state are pushed away from the trap by the push beam at 254nm. After the state selection, the long spectroscopic pulse is applied. At the end of the cycle, the MOT cooling beams and coils are turned on and the atoms are detected.}
	\label{figure3}
\end{figure}

The simple level structure of mercury allowed us to implement a simple form of state selection that is depicted in figure \ref{figure3} and described in the following.

We start with the lattice trapped atoms in the ground state equally distributed between $|^1S_0,m_F=1/2\rangle$ and $|^1S_0,m_F=-1/2\rangle$ states. The first step of the state selection is the excitation of a single Zeeman component  $|^1S_0,m_F\rangle\rightarrow|\phantom{}^3P_0,m_{F'}=m_F\rangle$, with a short $\pi$-pulse of the probe light in the presence of a magnetic field of $\sim0.1\,$mT. The second step is to apply a light pulse resonant with the cooling transition along the axis of the trap, which expels from the trap all the atoms remaining in $|^1S_0\rangle$. This way, only the $|^3P_0,m_{F'}=m_F\rangle$ state is left populated with a purity exceeding
$98\%$. We verified this by checking that no atoms are detected if we apply only the resonant pulse.  From this we conclude that the number of atoms remaining in the unwanted state is below our detection noise of 40 atoms \cite{Yi2011}, while the typical selected atom number is about $2\times 10^3$.

Using the simple state selection procedure described above, the clock transition is detected on a dark background, i.e. the signal comes only from $|^1S_0\rangle$ atoms which are populated by the interrogation probe pulse. This increases the signal to noise ratio of the frequency discrimination measurement as noise contributions from  atoms in the $|^1S_0,-m_F\rangle$ state are removed.

After the state selection sequence is completed, we perform Rabi spectroscopy on the $\pi$ component of the clock transition starting from the prepared $|^3P_0\rangle$ state by applying a $\pi-$pulse of the clock light and detecting atoms in the state $|^1S_0\rangle$ by collecting fluorescence emitted on the cooling transition $^1S_0\rightarrow\phantom{}^3P_1$. The fluorescence is induced with a molasses formed by four of the MOT beams and is digitized by the EM-CCD camera. Due to the geometry of our experiment, using four beams instead of six allows us to reduce the detection noise.

\subsection{Clock operation and short term stability}
During clock operation, we apply a standard procedure consisting in setting the probe frequency alternatively at $f_\mathrm{probe}+\Delta$ and $f_\mathrm{probe}-\Delta$, where $\Delta$ is roughly equal to the half linewidth, applying a $\pi$ Rabi pulse and correcting the probe frequency by a quantity proportional to the integrated difference between two subsequent fluorescence measurements. At equilibrium the probe frequency is thereby locked to the mercury transition frequency \cite{Riehle2005}.

In order to evaluate systematic effects, we use a technique of digital lock-in detection. Up to four independent integrators operating in the same way as the one described above are run in an interleaved manner. Each integrator runs for a period of few cycles and for each integrator we use a different value of the clock parameters set.
The frequency difference between the different integrator outputs yields information about the sensitivity of the clock transition frequency relatively to the modulated parameter(s) while the other parameters are kept constant. This allows us to measure different systematic frequency biases of the clock frequency while rejecting almost all the other perturbations as they are common mode for the different configurations.

\begin{figure}[htbp]
    \centering
	\includegraphics[width=0.65\textwidth]{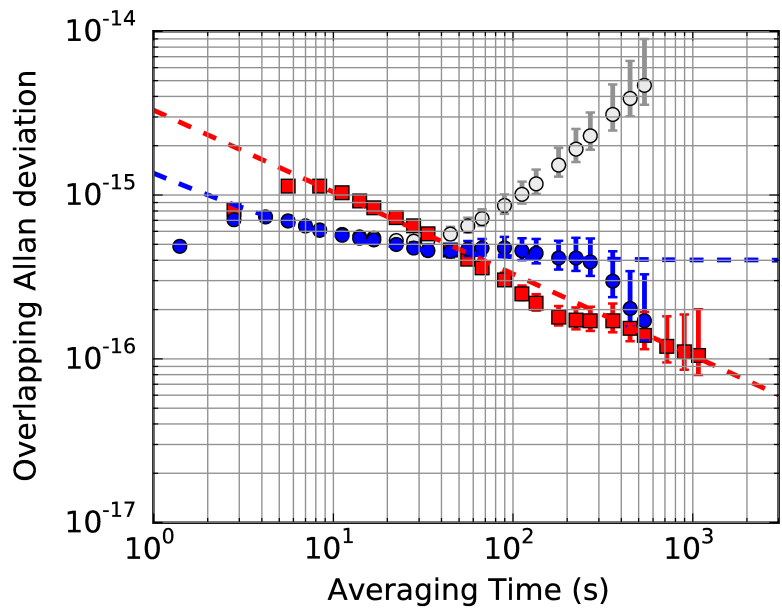}
	\caption{ Overlapping Allan standard deviation of the mercury clock center frequency for a typical measurement run (Rabi time of $0.1$~s, cycle time of $1.1$~s), when referenced to the high-finesse Fabry-Perot cavity. Gray dots represent the raw data, while the blue dots are corrected for the linear drift of the cavity. The blue dashed line is a model used to fit the dedrifted data : $\sigma_y(\tau)=\sqrt{\sigma^2_\mathrm{white}(\tau)+\sigma^2_\mathrm{flicker}(\tau)}$ where $\sigma_\mathrm{white}(\tau)= 1.3\times 10^{-15}\times\tau^{-1/2}$ represents white frequency  noise, and $\sigma_\mathrm{flicker}(\tau)=4\times 10^{-16}$ is the flicker floor of the ULE cavity.
	Red squares are overlapping Allan standard deviation of the frequency difference between the two Zeeman sublevels. This stability is representative of a differential measurement where the probe frequency is modulated every 2 cycles of the clock. The red dashed line corresponds here to a stability of $\sigma(\tau)= 3.4\times 10^{-15}\times\tau^{-1/2}$.}\label{fig:stability}
\end{figure}

However, due to cyclic operation of the clock with typically only $10\%$ of the time spent interrogating the atoms, frequency fluctuations of the clock laser cannot be completely eliminated but are aliased into the measurement bandwidth increasing the frequency noise and therefore limiting the short term stability of the clock (Dick effect, see \cite{Dick1990}). As during interleaved operation the cycle time to obtain a single measurement is increased, the effective instability of the interleaved measurement is increased with respect to normal clock operation.
This is illustrated in figure \ref{fig:stability} where we plot, for a typical clock configuration, the Allan standard deviation of the fractional frequency difference between the mercury atomic reference and the ultrastable light in raw form (gray) and with a linear drift removed (blue), and of the fractional frequency difference between the two $\pi$-components of the clock transition (see figure \ref{fig:MOLCfigure1}) with a sampling period of two clock cycles (red). The data in blue are fitted with a model that is the sum of two contributions: the white frequency noise of the mercury atomic reference for short averaging times and the flicker frequency noise of the cavity for long averaging times. The model is consistent with the expected performance of the ultrastable cavity (see section \ref{sec:clock}) and allows us to extract the short term stability of the mercury clock as $1.2 \times 10^{-15}$. For the differential measurement, in spite of the slightly degraded stability, the measurement averages down to a $1\times 10^{-16}$ statistical uncertainty in about $700\,$s, allowing for measurements below the flicker floor of the cavity.

%% file: systematics.tex
In this section we describe the evaluation of the physical perturbations affecting the clock frequency measurement. This evaluation is paramount in order to be able to evaluate the frequency comparisons against other frequency standards presented in the next section.

The main perturbations affecting the mercury clock frequency measurement are those related to the trap AC Stark shift, the external magnetic field, the thermal radiation (black body radiation shift, BBR), the atomic density in the trap, and the pulsed interrogation.

\subsection{Lattice light shift}\label{sec:lattice_shift}
We first consider the effect of the trapping light on the clock transition.
Trapping the atoms in a lattice allows for cancellation of Doppler and other motional effects, but it has the consequence of introducing an AC Stark shift on the clock levels, and in turn a frequency shift of the clock transition. As first pointed out in \cite{Katori2003}, tuning the frequency of the trapping light to the so-called ``magic frequency'' \numagic\  for which the scalar polarizabilities of the clock states are equal allows for a cancellation to first order of this frequency shift. Furthermore as $^{199}$Hg has spin 1/2, the tensor component of differential polarizabilities are zero by symmetry.

Given the relatively small trap depth available in our experiment (see section \ref{sec:lattice}) and the predicted value of non-linear terms for mercury \cite{Katori2015}, we expect that non-linear lattice shifts will not be resolved at our present resolution. We therefore analyzed the data according to a linear model of the lattice light shift (see \cite{Katori2015}):
\begin{equation}
    \Delta\nu_\mathrm{LLS} = \frac{1}{h \alpha^s_{^1S_0}} \left. \frac{\partial \Delta\alpha^s}{\partial \nu}\right\vert_{\nu_l}\,(\nu_l - \numagic)\, U_l = \left. \frac{\partial \Delta\kappa^s}{\partial \nu}\right\vert_{\nu_l}\,(\nu_l - \numagic)\, U_l,
\end{equation}
where $\Delta\alpha^s(\nu) = \alpha^s_{^3P_0}(\nu) - \alpha^s_{^1S_0}(\nu)$ is the difference between the values of scalar polarizabilities of the ground and excited states of the clock at the operating frequency of the lattice $\nu_l$, and $U_l$ is the operating depth of the lattice.
\begin{figure}[htbp]
    \centering
    \includegraphics[width=10.16cm]{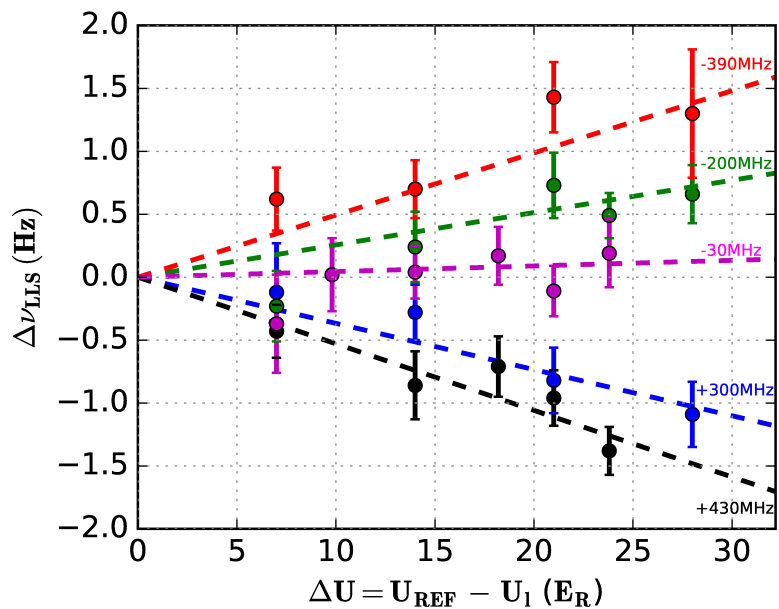}
    \caption{Lattice AC stark shift (Hz) as a function of the differential lattice depth in recoil energy. The detuning (in MHz) from 826~855~533 MHz corresponding to each curve is shown on the plot with matching color. The dotted lines show the result of a fit of the whole dataset with a linear model }
    \label{fig:systematics_figure1}
\end{figure}

We study this effect using digital lock-in detection having two integrators running on each one of the $\pi$-components of the clock transition at the nominal depth $U_\mathrm{REF}=56 \erec$ and a second pair of integrators running at a different depth $U_l$ between $50$ and $25\erec$. The linear combination
$$\Delta\nu = 1/2\,\left[ \nu_m(\pi^+,U_\mathrm{REF}) + \nu_m(\pi^-,U_\mathrm{REF}) - \nu_m(\pi^+,U_l) - \nu_m(\pi^-,U_l) \right],$$
where $\nu_m(\pi,U)$ is the clock frequency measured by the integrator operating on the $\pi$ component at the depth $U$, is a measure of $\Delta\nu_\mathrm{LLS}(\nu_l,U_\mathrm{REF}) - \Delta\nu_\mathrm{LLS}(\nu_l,U_l)$. We note that the measured $\Delta \nu$ is insensitive to lattice vector shift, and to magnetic field instabilities to first order (see section \ref{sec:zeeman}).

By repeating these integrations for different lattice frequencies, we obtain the dataset plotted in figure \ref{fig:systematics_figure1}, as a function of the difference $U_\mathrm{REF} - U_l$. In the figure we also plot a fit of the whole dataset with the function: $\Delta \nu =a (\nu_l - \numagic) (U_{REF}-U_l)$ with free parameters $a$ and \numagic.

From the fit we extract  the values $\partial \Delta \kappa^s/\partial \nu(\numagic) = a =-1.25(7) \times 10^{-4}\,$Hz/\erec/MHz and an estimation of the magic frequency $\numagic = 826\,855\,539(21)\,$MHz, an improvement by more than two orders of magnitude with respect to our previous determination \cite{McFerran2012} and in agreement with the measurement from \cite{Yamanaka2015}. A $\chi^2$ test yields a probability of 0.96 for the fit, which gives us confidence that higher order terms are indeed not resolved at our present level of uncertainty.

In order to support this claim and to take at least partially into account the non-linear terms, we evaluated the lattice light shift with the complete model presented in reference \cite{Katori2015} using the theoretical estimation of coefficients of the non-linear terms and our experimental determination of the linear ones. From this non-linear model we generate datapoints at the values of detuning and lattice depth probed experimentally and perform a linear analysis in the same way as we do for the experimental data. We check then that the values of $a$ and \numagic\ obtained are consistent with the experimental ones.
From the measured coefficients, evaluating our linear model at our nominal trap depth of 56 \erec\ and taking into account the uncertainty of the wavemeter used to lock the trapping laser to the magic wavelength (see section \ref{sec:lattice}), we infer a lattice light shift of $4\times 10^{-17}$ with an uncertainty of $1.38\times 10^{-16}$.  Using the full model of ref. \cite{Katori2015}, as mentioned above, we evaluate the contribution of non-linear terms as $-6\times \tm{17}$ with an uncertainty of $3.6\times\tm{17}$ including a (conservative) $50\%$ relative uncertainty on the theoretical non-linear coefficients and $15\,\mu$K uncertainty on the atomic temperature in the lattice.

In the following we therefore apply a typical relative correction for lattice light shift of $2\times \tm{17}$ with a total uncertainty of $1.43\times \tm{16}$, dominated by the linear term. This is the largest contribution to our total uncertainty budget.

\subsection{Blackbody radiation shift (BBR)}
The black body radiation (BBR) shift is due to differential AC Stark shift of the clock levels due to the electromagnetic radiation in thermal equilibrium with the environment of the atomic sample.
The frequency shift induced by the environment at a temperature $T$ can be written as
\begin{equation}
    \delta\nu_{BBR}=-\frac{2\sigma}{h \epsilon_0 c} \Delta \alpha_\mathrm{st} T^4(1+\eta T^2)
    \end{equation}
where $\sigma$ is the Stefan-Boltzmann constant, $\Delta \alpha_\mathrm{st}$ is the static differential polarizability between the excited and the ground state, and $\eta$ is the so-called dynamical coefficient. Based on the frequencies of the relevant transitions we expect that Hg should behave similarly to Yb for which $\eta T^2 < 0.02$ \cite{Beloy2014}. The resulting frequency shift and associated uncertainty are therefore expected to be below $1\times\tm{17}$ and therefore negligible with respect to the present uncertainty over the static term.

As BBR shift is the most significant item in the uncertainty budget of the best optical lattice clocks based on Sr \cite{Nicholson2015}, \cite{Falke2014}, \cite{LeTargat2013} or Yb \cite{Lemke2009} atoms, much effort has been made in tackling this effect in the literature. This includes ultra-precise radiation thermometry \cite{Bloom2014}, radiation shielding \cite{Beloy2014}, or cryogenic setups \cite{Ushijima2015}.

In the case of Hg, the sensitivity to this effect is much smaller than for Sr (30 times) or Yb (16 times), and therefore controlling this shift down to the $10^{-17}$ level of uncertainty only requires the knowledge of the thermal environment of the atoms at the level of a few kelvin. In order to evaluate both temperature and thermal gradients in the environment seen by the atoms, we have put three Pt100 sensors in contact with the vacuum chamber. These sensors show good agreement, with measured temperature differences not higher than 1K . Assuming a $10\%$ uncertainty on the differential polarizability \cite{Hachisu2008} and with a measured temperature around the atoms of $(301.0 \pm 1.5)\,$K, we get a shift of $-1.6 \times 10^{-16}$ with an uncertainty of $2\times 10^{-17}$.

\subsection{Collision shift}
Collision shift is a prominent contribution to many atomic clocks uncertainty budgets. It is the main limitation to the accuracy of the best Cs \cite{Guena2012} and Rb \cite{Guena2014} microwave frequency standards and has been observed in both Sr \cite{Swallows2011} and Yb \cite{Lemke2009} optical lattice clocks.

We have used two different methods to modulate the probed atom number between a high density (HD) and a low density (LD) configuration in order to evaluate the collisional shift.
\begin{figure}[htbp]
    \centering
	\subfloat[]{
	\includegraphics[width=7cm]{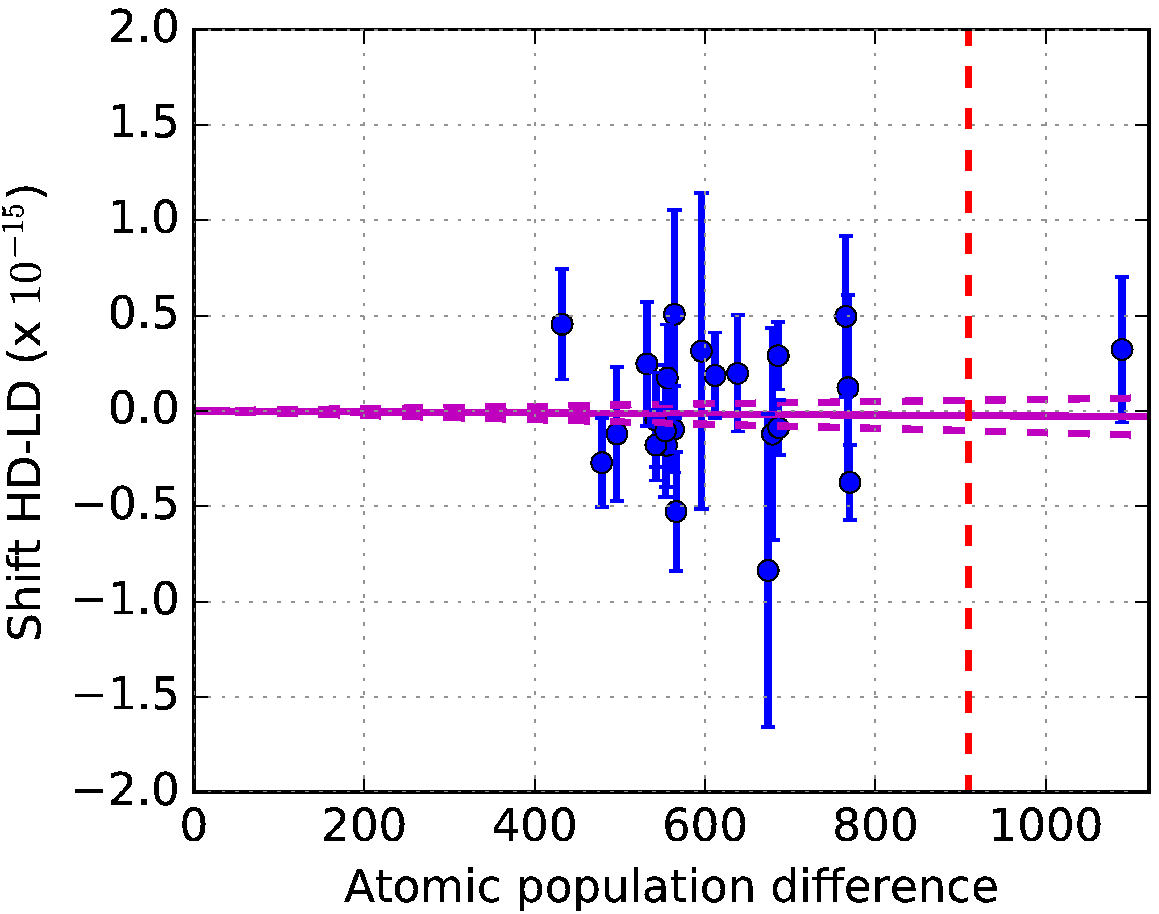}
	\label{fig:systematics_figure2a}}
    \hspace{0.2cm}
    \subfloat[]{
    \includegraphics[width=7cm]{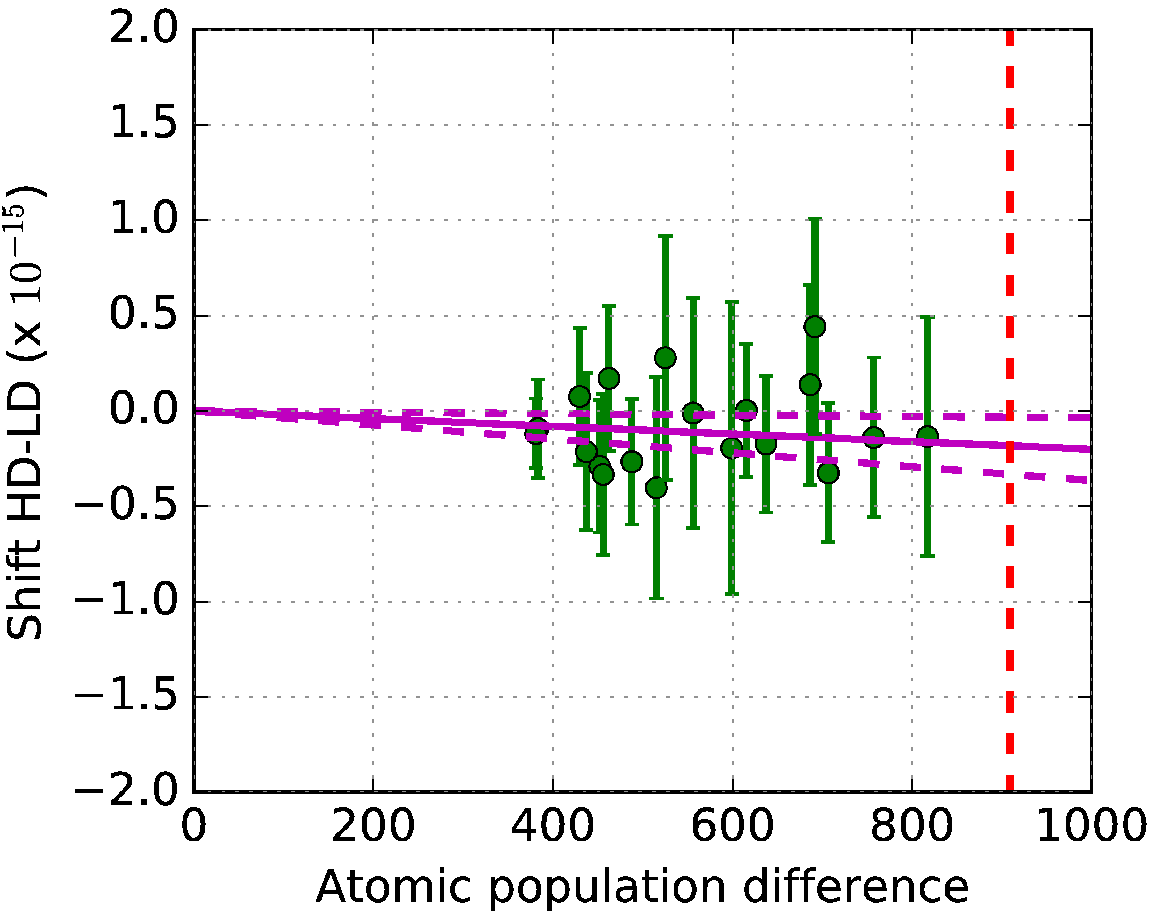}
    \label{fig:systematics_figure2b}}
    \caption{Differential collisional shift shown in relative units as a function of the population difference for
    \protect\subref{fig:systematics_figure2a} the state selection time variation method and
    \protect\subref{fig:systematics_figure2b} the MOT loading time variation method. The data sets are fitted with straight lines with no offset, since the differential method must give zero shift at zero modulation of the parameters. The 1$\sigma$ confidence interval is given by the magenta dotted lines on each curve. The vertical red dashed line corresponds to the number of atoms probed during the nominal operation of the clock and allows for a graphical estimation of the collision shift.}
    \label{fig:systematics_figure2}
\end{figure}

In a first series of measurements, we modulate the duration of the state selection pulse duration (see section \ref{sec:state_selection}), while keeping the power constant, thereby changing the fraction of atoms excited to the $^3P_0$ state. The results are presented on figure \ref{fig:systematics_figure2a} and summarized in table \ref{tab:collisional_shift}. The extrapolated shift for the nominal atom number is $(-1.6 \pm 1.4)\times 10^{-16}$.
In order to check for the presence of a bias in this technique such as a phase transient in the AOM used to shape the state selection pulse or a non-linear change in the density distribution with the atom number, we perform a second series of measurement by varying the loading time of the MOT.

The results are shown in figure \ref{fig:systematics_figure2b} and summarized in table \ref{tab:collisional_shift}. The extrapolated shift for the employed atom numbers is $-2(7)\times 10^{-17}$.
\begin{table}[hb]
\caption{\label{tab:collisional_shift}Summary of collisional shift measurements}
\begin{indented}
\item[]\begin{tabular}[top]{@{}l|lll}
\br
 Method: &Loading time&State selection pulse&Combined\\
\mr
Slope ($10^{-20}$/atom)&-2.6&-20&-6.5\\
1$\sigma$ uncertainty ($10^{-20}$/atom)&8.8&16&7.8\\
\mr
Shift ($10^{-17}$)&-2.1&-16.2&-5.2\\
1$\sigma$ uncertainty ($10^{-17}$)&7.1&14.0&6.4\\
\br
\end{tabular}
\end{indented}
\end{table}

Table \ref{tab:collisional_shift} sums up the results for the two methods, extrapolating the results to our nominal operating atom number of $\simeq$ 900 atoms probed in the lattice. As we do not have any reason to favour one method over the other, we combine all the data treating it as a single dataset. Fitting with a linear function, we find a shift of $-5.2(6.4) \times 10^{-17}$, compatible with zero within a 1$\sigma$ uncertainty. The results of this procedure are shown in the third column of table \ref{tab:collisional_shift}.

Another source of systematic uncertainty related to atom-atom interactions are background collisions. We evaluate this shift by applying the theory of ref. \cite{Gibble2013} using the measured fraction of detected atoms and mercury properties from refs. \cite{Mitroy2010,Kullie2014}. We obtain a shift of $-2 \times 10^{-17}$. However as the experimental configuration and the interrogation method used in this work are not fully matching the theoretical treatment of \cite{Gibble2013}, we decide not to apply any correction and include background collisions in our uncertainty budget with a $150\%$ uncertainty.

\subsection{Zeeman shifts}\label{sec:zeeman}
The effect of an external magnetic field on the clock transition is to introduce a frequency shift on the $\pi$ components given by
$$\Delta\nu_{mag} = m_F g_F \mu_B B_z - \beta_Z B_z^2$$
where $g_F$ is the $g$-factor of the transition, $\mu_B$ is the Bohr's magneton in frequency units, $\beta_Z$ is the $2^{nd}$ order Zeeman coefficient and $B_z$ is the projection of the field along the polarization of the probe beam.
Using spectroscopic measurements we have mapped and subsequently zeroed the magnetic field at the atom site with a resolution of 3 $\mu T$ in all three directions of space.
We then apply a quantization magnetic field $B_z \simeq 280\,\mu$T parallel to the probe polarization to split the two $\pi$ Zeeman components of the transition.

Similarly to the lattice light shift case (see section \ref{sec:lattice_shift}) we perform digital lock-in detection running four independent integrators, one pair operating at a reference bias field $B_\mathrm{REF}$ and locking on each of the two $\pi$ components, and another pair operating at a variable bias field $B_\mathrm{bias}$ also on the two $\pi$ components of the transition.
For each integrator pair we then construct the two quantities:
\begin{eqnarray}
        \nu_m(B) &=& 1/2 (\nu_m(\pi^+,B) + \nu_m(\pi^-,B)) \nonumber\\
        \Delta \nu(B) &=& \nu_m(\pi^+,B) - \nu_m(\pi^-,B)
\end{eqnarray}
which are related to the relevant physical parameters by:
\begin{eqnarray}
        \nu_m(B_\mathrm{bias}) - \nu_m(B_\mathrm{REF}) &=& - \beta_Z (B_\mathrm{bias}^2 - B_\mathrm{REF}^2)\nonumber\\
        \Delta \nu(B) &=& g_F \mu_B B + \Delta_{VS},
\end{eqnarray}
where $\Delta_{VS}$ is the vector component of the lattice light shift.

A measurement of this last quantity can be obtained by analyzing the data presented in section \ref{sec:lattice_shift} extracting the splitting between the two $\pi$ components as a function of the lattice depth. At the nominal trap depth we obtain $\Delta_{VS} \simeq 15\mathrm{Hz} \ll \Delta \nu(B_\mathrm{bias})$, indicating a residual ellipticity of the probe beam of about $3\%$ if we use the vector polarizability given in \cite{Hachisu2008}.

\begin{figure}[htbp]
    \centering
	\includegraphics[width=.6\textwidth]{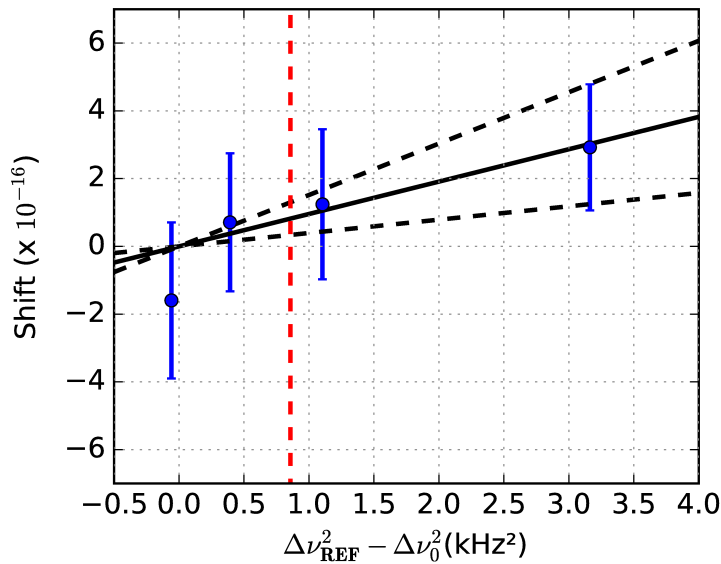}
    \caption{Quadratic Zeeman shift as a function of the square of the splitting between the two $\pi$ components. The plain black line is a linear fit to the data, and the two black dotted lines represent 1$\sigma$ confidence intervals on the fit. The red dotted line is our nominal operating point.}
    \label{fig:systematics_figure3}
\end{figure}

Neglecting the vector light shift, we can use the measurement of $\Delta \nu(B)$ as an \emph{in situ} calibration of the magnetic field and we get the quadratic Zeeman shift
\begin{equation}\label{eq:qzs}
    \Delta\nu_\mathrm{QZS} = \frac{\beta_Z}{(g_F \mu_B)^2} (\Delta\nu(B_{REF})^2-\Delta \nu(B_\mathrm{bias})^2).
\end{equation}
We plot in figure \ref{fig:systematics_figure3} the differential shift of the center frequency of the clock transition as a function of the difference of the square of the splittings for the two interleaved configurations.

From the slope of a fit to the data according to eq. \ref{eq:qzs}, we obtain the atomic coefficient $\gamma_Z=\beta_Z (g_F \mu B)^{-2}$. We find $\gamma_Z=1.1(6)\times \tm{7}\,\mathrm{Hz}^{-1}$, resulting in a frequency shift of  $-8.2\times \tm{17}$ at our nominal splitting of $926\,$Hz with an uncertainty of $4.8\times 10^{-17}$, limited by the statistics of the measurement.

\subsection{Clock pulse frequency chirp}
\begin{figure}[htbp]
    \centering
    \begin{tabular}{c}
    \subfloat[]{
	\includegraphics[width=14.16cm]{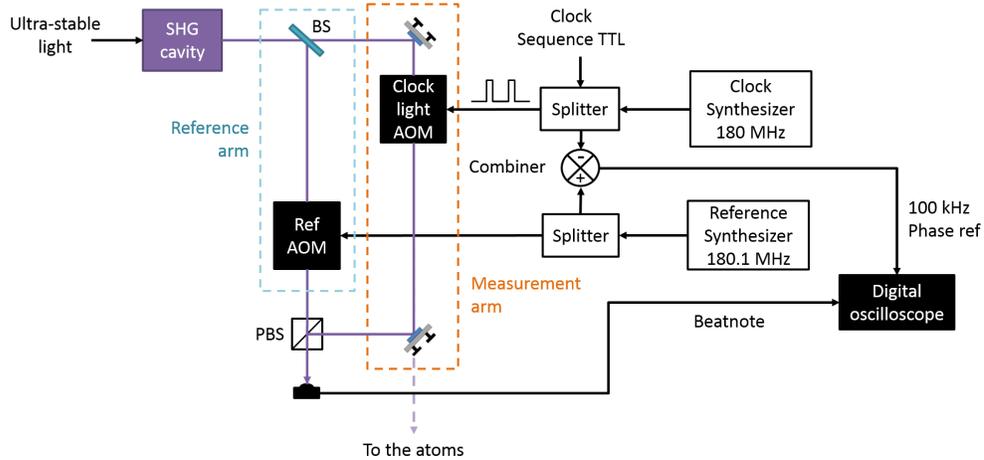}
	\label{fig:systematics_figure4a_v1}}\\
	\subfloat[]{
	\includegraphics[width=6cm]{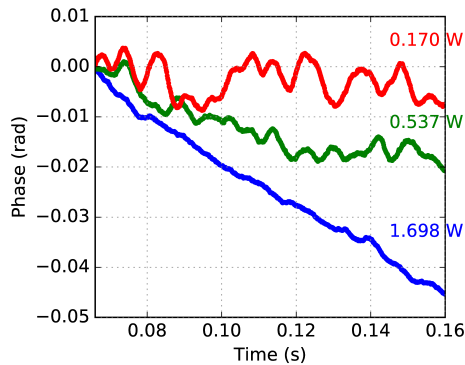}
	\label{fig:systematics_figure4b_v1}}
    \hspace{0.2cm}
    \subfloat[]{
    \includegraphics[width=6cm]{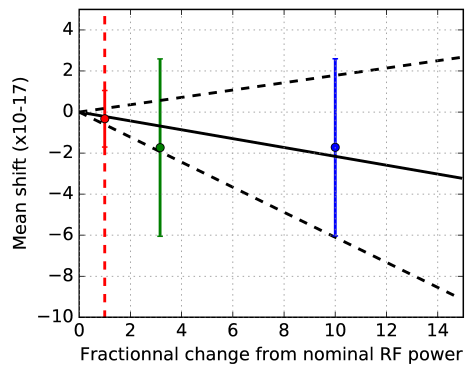}
    \label{fig:systematics_figure4c_v1}}
    \end{tabular} \hfill
    \caption{
    \protect\subref{fig:systematics_figure4a_v1} Scheme of the AOM phase-transient measurement setup. Two synthesizers are used to drive the clock and the reference arm AOMs, respectively at 180 and 180.1 MHz. The  pulses are created via a TTL signal sent to the clock AOM. We take a small amount of clock light at 266 nm from the output of the second harmonic generation cavity using a beam-sampler (BS). This light is sent to the reference AOM (Ref AOM), while the remaining light goes through the usual clock beam path and is pulsed through the clock light AOM. The measurement and reference arms are overlapped at the output port of a polarizing beam-splitter (PBS) and sent to a photodiode. The resulting beatnote at $100\,$kHz is acquired with a digital oscilloscope and the phase is extracted by numerically computing the amplitude of the in-phase and in-quadrature components. Two splitters and a mixer provide the phase reference for the demodulation. \protect\subref{fig:systematics_figure4b_v1} Typical phase of the beatnote inside the probe pulses time window for several RF powers.
   	\protect\subref{fig:systematics_figure4c_v1} Frequency shift introduced by the AOM phase chirp as a function of the fractionnal RF power with respect to our nominal RF power of 170 mW. The black dotted lines materialize 1$\sigma$ confidence interval on the fit represented in full black line, and the nominal operating point of the clock is indicated by the red dotted line.}
    \label{fig:systematics_figure4}
\end{figure}

The clock pulses are created using an AOM driven at the frequency of 180 MHz (see Figs. \ref{fig:MOLCfigure1} and \ref{fig:systematics_figure4a_v1}).
The RF power dissipated in the AOM can cause phase transients in the AOM, in turn resulting in a frequency shift \cite{Kazda2015}. We realize an interferometric measurement of this phase transient by analyzing in the time domain the beatnote between the interrogation pulses and a CW beam passing through another identical AOM. A schematic of the measurement setup is shown in figure \ref{fig:systematics_figure4a_v1}.

Typical results of the reconstructed phase transient are shown in figure \ref{fig:systematics_figure4b_v1} for several RF driving powers of the AOM, averaged over 654 cycles of the clock. Taking the slope from this plot and averaging over several realizations we can evaluate the frequency shift associated with the phase transients of the probe light $\Delta\nu_{probe}=\frac{1}{2\pi}\frac{d\phi}{dt}$, which is plotted in Figure \ref{fig:systematics_figure4c_v1} with matching colors.
For $170\,$mW of RF power driving the AOM, which corresponds to the typical value for the operation of the clock at $\sim 0.1\,$s of Rabi time, we obtain a shift consistent with zero with an uncertainty below \tm{17}.

\subsection{Final uncertainty budget}
A summary of the evaluated frequency shifts on the mercury $^1S_0\rightarrow\phantom{}^3P_0$ transition and of the associated corrections which we apply to get the unperturbed frequency of the clock is given in table \ref{tab:uncertainty}. For the sake of completeness we include the estimate of the probe light shift (in the $10^{-18}$ range) based on the measured Rabi frequency of the clock transition and the contributions to the polarisability of the main allowed transitions for the two clock levels.
Summing uncertainty contributions quadratically, we get the total fractional uncertainty on the clock transition frequency as $1.7\times 10^{-16}$, a factor 30 improvement over our last reported uncertainty \cite{McFerran2012} and only a factor of 2 higher than the recent evaluation of a similar system \cite{Yamanaka2015}.
\begin{table}[htbp]
\caption{Uncertainty budget}\label{tab:uncertainty}
\begin{indented}
\centering
\item[]\begin{tabular}[top]{@{}lll}
\br
Effect&Correction ($10^{-17}$)&Uncertainty($10^{-17}$)\\
\mr
Second order Zeeman&8.2&4.8\\
Cold collisions & 5.2 & 6.4\\
Background gas collisions & 0& 3.0\\
Lattice light shift (linear+non-linear)& 2.2 & 14.3\\
BBR&16.1&2.2\\
Probe light-shift&0&0.1\\
AOM Chirp&0.2&0.4\\
\mr
Total&25.5& 16.8\\
\br
\end{tabular}
\end{indented}
\end{table}

%% file: comparison.tex
In this section we report the results of a frequency comparison between the mercury optical lattice clock described in the previous section and three different frequency standards. A primary frequency standard, an atomic fountain based on Cs (FO2-Cs) with an accuracy of $2.4\times\tm{16}$ \cite{Guena2012}, and two secondary standards, one atomic fountain operating on the microwave ground state hyperfine splitting in $^{87}$Rb at $6.8\,$GHz (FO2-Rb, uncertainty $2.9\times\tm{16}$ \cite{Guena2014}) and a lattice clock operating on the visible $^1S_0 \rightarrow \phantom{}^3P_0$ transition in $\phantom{}^{87}$Sr at $429\,$THz (Sr2, uncertainty $4.1\times\tm{17}$ \cite{Lodewyck2016}).

\begin{figure}[ht]
	\centering
	\includegraphics[width=.95\textwidth]{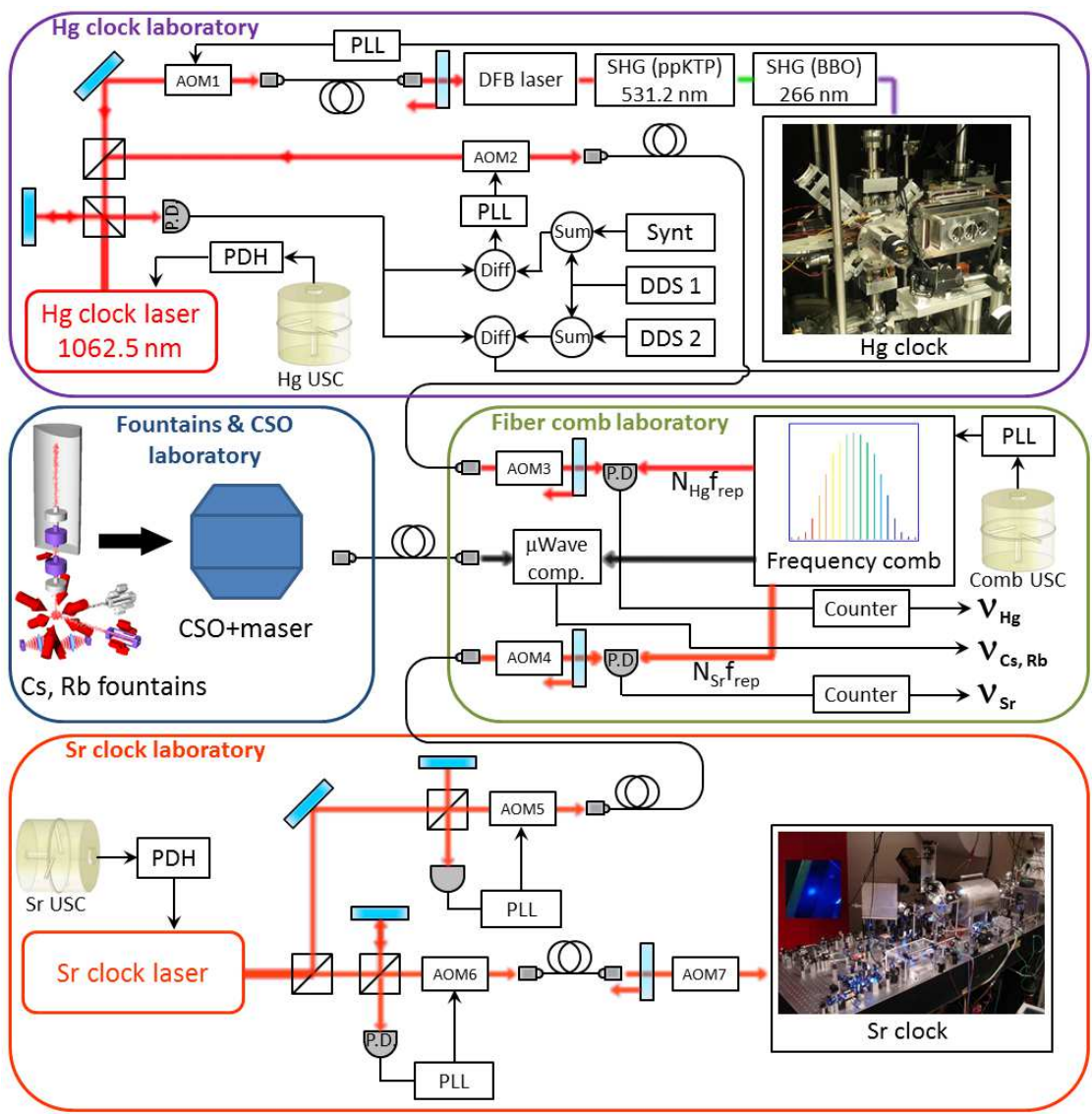}
	\caption{Mercury and strontium optical lattice clocks frequency comparison scheme. Part of the probe light of each clock is sent to the fiber comb via frequency stabilized optical links. For the mercury clock the stabilized path does not include the two stages of frequency doubling.
	The fiber comb is also connected to the maser+CSO ensemble via a frequency stabilized link, locked to the Cs and Rb fountains. The comb simultaneously measures all the frequencies. Notice that the comb measures the frequency of the Hg clock in the infra-red and not in the UV. The Hg and Sr clocks were not synchronized during the comparison.}
	\label{fig:Comparison_Structure_Figure1}
\end{figure}

The comparison setup is sketched in figure \ref{fig:Comparison_Structure_Figure1}. The optical clocks are located in two different laboratories in the same building. A fraction of the probe light of each clock is sent via optical fibers to a fiber comb located in a third laboratory. At the end of both links a small amount of light is reflected and coupled back into the fibers. The reflected light is beat with the input light and the beatnote is used for fiber noise cancellation. The fiber comb also has a frequency stabilized link to the H-maser+cryogenic sapphire oscillator (CSO) ensemble. The beatnote of the clock light and an appropriate comb tooth for each clock is measured and counted in the fiber comb laboratory against the H-maser reference. As the beatnotes for the two optical clocks are counted simultaneously it is possible to compute the frequency ratio in a way that rejects the H-maser noise to better than \tm{20}.

Data of the simultaneous measurements of the H-maser frequency against the Cs and Rb fountains was provided by the fountain laboratory. This allowed us to perform a frequency comparison of our Hg clock with the microwave frequency standards.

The results of the comparison presented below take into account the gravitational redshift due to a height difference of $4.25(2)\,$m ($4.17(2)\,$m) between the Hg clock and the Cs (Rb) fountain, calculated based on a leveling campaign carried out in 2013. The gravitational red shifts introduced by the height difference of $3(1)\,$cm between the Hg and Sr clocks (Hg lying higher than Sr) contribute to less than \tm{17}.

\subsection{Mercury versus microwave standards}
\begin{figure}[htb]
    \centering
	\includegraphics[width=.9\textwidth]{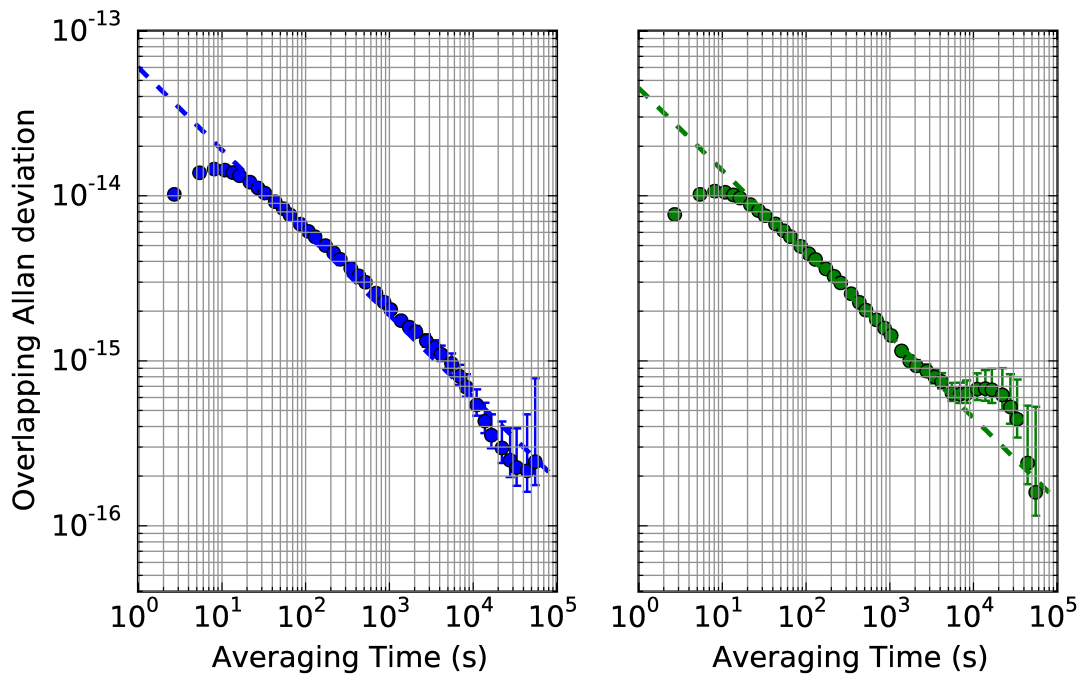}
    \caption{Overlapping Allan deviation of the frequency ratio between (left) Hg and FO2-Cs and (right) Hg and FO2-Rb. The dotted lines represent the stabilities of the fountains (limited by the quantum projection noise) over the length of the measurement campaign, averaging as $1/\sqrt{\tau}$.}\label{fig:comp_uwave}
\end{figure}

Comparison data in the microwave domain are shown in figure \ref{fig:comp_uwave} where we plot the overlapping Allan standard deviation of the measured frequency ratio for $^{133}$Cs and $^{87}$Rb respectively.

The observed frequency stabilities are compatible with white frequency noise at $6\times 10^{-14} \tau^{-1/2}$ for Cs, and $4.5\times 10^{-14} \tau^{-1/2}$ for Rb, fully limited by the short-term stability of the fountains. By considering a total measurement duration of about $39.4\,$hours over 10 days, limited by the up-time of the mercury clock, we obtain statistical uncertainties of $1.7\times 10^{-16}$ and $1.3\times 10^{-16}$ respectively for Cs and Rb frequency comparison.

\begin{figure}[htb]
    \centering
	\subfloat[]{
		\includegraphics[width=7cm]{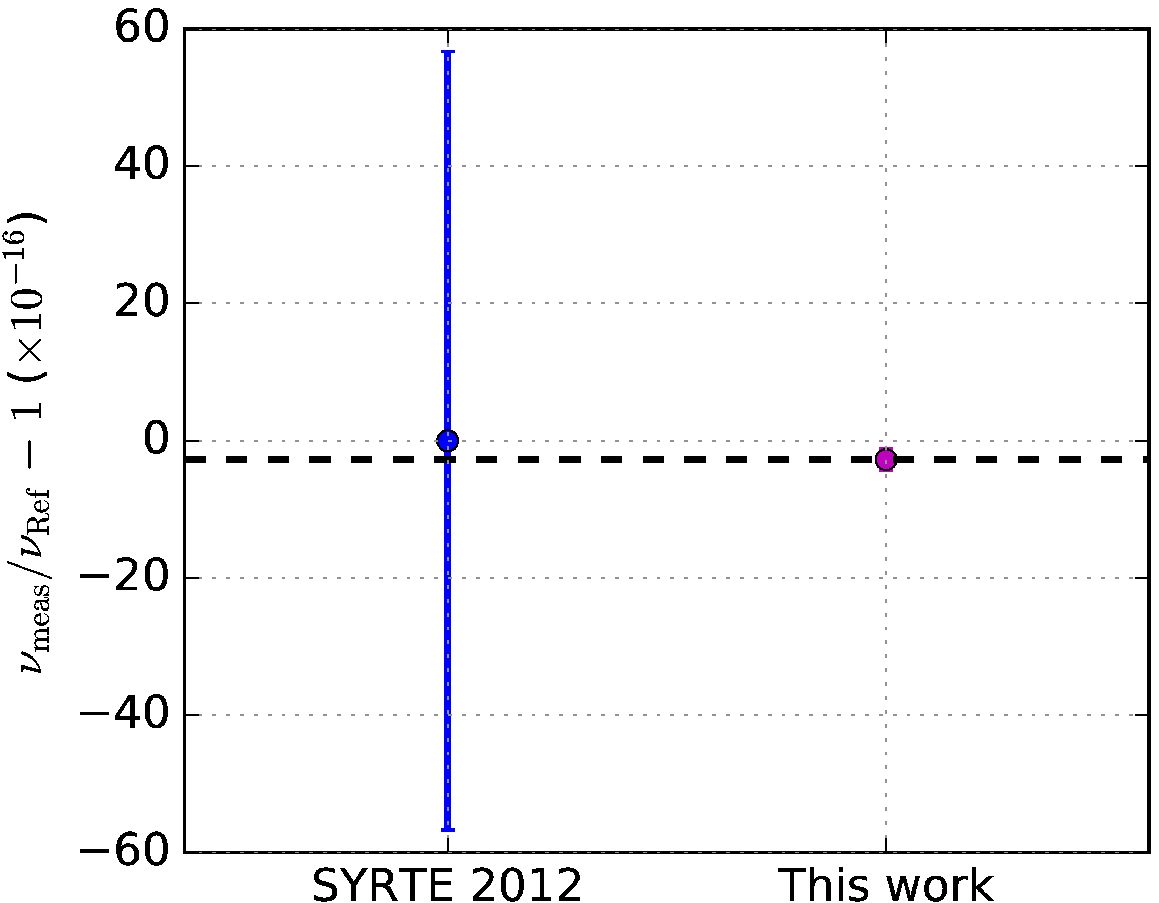}
		\label{fig:Comp_results_figure3a_v1}}
    \hspace{0.2cm}
    \subfloat[]{
    	\includegraphics[width=7cm]{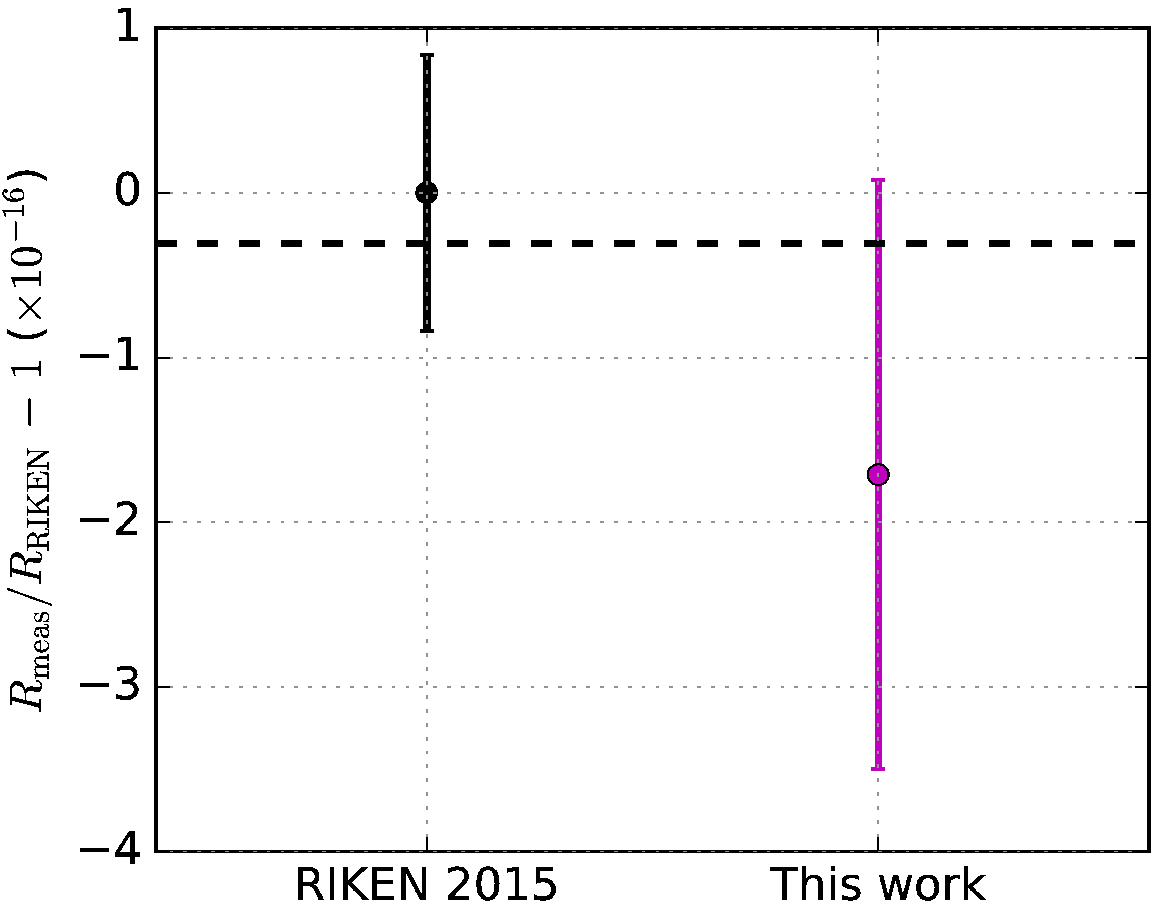}
		\label{fig:Comp_results_figure3b_v1}}
    \caption{
    \protect\subref{fig:Comp_results_figure3a_v1} Ratio between the measured absolute frequency of the mercury clock transition and the value of SYRTE's measurement in 2012 taken as reference (offset by unity). The 2012 measurement is in blue and the present measurement in violet. The horizontal black dotted line is the weighted mean of the two values.
    \protect\subref{fig:Comp_results_figure3b_v1} Ratio between the measured $\nu_{\mathrm{Hg}}/\nu_{\mathrm{Sr}}$ and the value from RIKEN group \cite{Yamanaka2015} (offset by unity). We have plotted here the measurement in reference \cite{Yamanaka2015} (black) and the measurement of this work (violet). The horizontal black dotted line is the weighted mean of the two values.}
\end{figure}

From the Hg/Cs frequency comparison we obtain the value of the frequency ratio 122~769.552~729~311~011(45),
 where the total uncertainty is the sum of three contributions: the systematic uncertainty on the mercury clock, the accuracy of the $^{133}$Cs fountain and of the frequency comb and the statistical uncertainty over the ratio measurement. By definition of the SI second, the Hg/Cs ratio gives us the value of the absolute frequency of the $^1S_0\rightarrow\phantom{}^3P_0$ transition in neutral mercury as $\nu_{\mathrm{Hg}} = 1 128\,575\,290\,808\,154.62\,$Hz $\pm0.19\,$Hz (statistical) $\pm0.38\,$Hz (systematic including Cs accuracy).
As can be seen in Figure \ref{fig:Comp_results_figure3a_v1}, this value is in excellent agreement with the previous value measured in our laboratory in 2012 \cite{McFerran2012,McFerran2015} and the accuracy is improved by roughly a factor of 20.

Our Hg/Rb comparison is the first direct measurement of the ratio of the two clock frequencies. We obtain the value $165\,124.754\,879\,997\,258(62)$ , where the quoted relative uncertainty of $3.7\times\tm{16}$ includes both statistical and systematic uncertainties. Using the measured Hg/Rb ratio we can obtain another absolute frequency measurement of the frequency of the mercury transition by noting that
$\nu_\mathrm{Hg} = (\nu_\mathrm{Hg}/\nu_\mathrm{Rb}) \times (\nu_\mathrm{Rb}/\nu_\mathrm{Cs}) \nu_\mathrm{Cs}$. Using the best known experimental value for the Rb/Cs ratio \cite{Guena2014} which provided the recommended value of the Rb secondary representation of the second \cite{BIPM_LOR}, we obtain $\nu_{\mathrm{Hg}} =$ 1~128~575~290~808~154.19~Hz $\pm0.15\,$Hz (stat) $\pm0.40\,$Hz (syst).

\subsection{Mercury versus strontium optical frequency ratio}
\begin{figure}[h]
    \centering
	\includegraphics[width=.95\textwidth]{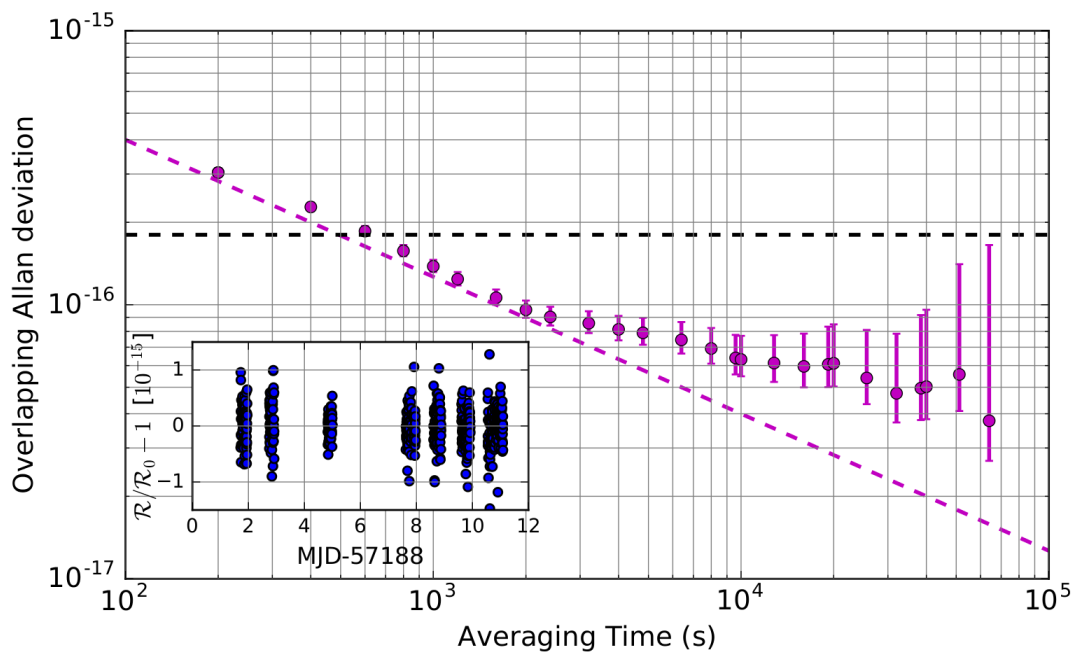}
	\caption{Overlapping Allan deviation of the frequency ratio between Hg and Sr. The dotted magenta line represents $4\times 10^{-15}$/$\sqrt{\tau}$ and the dotted black line the accuracy of the mercury clock as evaluated in section \ref{sec:systematics}. In the inset we plot the time series of the
	frequency comparison measurement. Each dark blue dot represents the relative deviation from the average of the full data set ($\mathcal{R}_0$), where $\mathcal{R}$ is the average of 200 consecutive clock cycles.}\label{fig:hgsrtime}
\end{figure}

The results of the $\nu_\mathrm{Hg}$/$\nu_{\mathrm{Sr}}$ direct optical-to-optical frequency comparison via the frequency comb are shown in the inset of Figure \ref{fig:hgsrtime} averaged over $200\,$s wide bins. The total overlapped up-time of the two clocks is $36.6\,$hours. The Allan deviation of these data is shown in Figure \ref{fig:hgsrtime}.
As expected the stability is much better than in the microwave to optical comparison and is consistent with white frequency noise at the level of $4\times \tm{15}$ at 1 s up to roughly 2 hours, and then with frequency flicker noise around $5\times 10^{-17}$, below the present accuracy of the Hg lattice clock.

We obtain the value $\nu_{\mathrm{Hg}}/\nu_{\mathrm{Sr}} = 2.629\,314\,209\,898 \,909\,15(46)$. The quoted uncertainty ($1.8\times\tm{16}$ in relative units) is the sum of statistical and systematic uncertainties and is dominated by the systematics of the mercury lattice clock as evaluated in section \ref{sec:systematics} since the Sr clock contributes at the level of only $4.1\times\tm{17}$. The obtained value is in good agreement with the value measured in \cite{Yamanaka2015}.

\section{Conclusions}
The three clock frequency ratios measured in this work are listed in table \ref{Frequency ratio measurements}. The $\nu_{\mathrm{Hg}}/\nu_{\mathrm{Rb}}$ frequency ratio is reported for the first time. To our knowledge the $\nu_{\mathrm{Hg}}/\nu_{\mathrm{Sr}}$ frequency ratio is now one of the best known physical quantities measured independently in different laboratories,
on-par with the $\nu_{\mathrm{Sr}}/\nu_{\mathrm{Cs}}$ ratio measured both at SYRTE \cite{LeTargat2013} and at PTB \cite{Falke2014} with an accuracy in the low $10^{-16}$ range and a factor of two better than the $\nu_{\mathrm{Yb}^+,E3}/\nu_{\mathrm{Yb}^+,E2}$ ratio measured at NPL \cite{Godun2014} and at PTB \cite{Huntemann2014}.
\begin{table}[ht]
\caption{\label{Frequency ratio measurements}Frequency ratio measurements}
\begin{indented}
\centering
\item[]\begin{tabular}[top]{@{}llc}
\br
Ratio&Measured value&Fract. unc. $\times 10^{-16}$ \\
\mr
$\nu_{\mathrm{Hg}}/\nu_{\mathrm{Cs}}$ & 122 769.552 729 311 011 (45) & 3.7 \\
$\nu_{\mathrm{Hg}}/\nu_{\mathrm{Rb}}$ & 165 124.754 879 997 258 (62) & 3.8 \\
$\nu_{\mathrm{Hg}}/\nu_{\mathrm{Sr}}$ & 2.629 314 209 898 909 15 (46) & 1.7 \\
\br
\end{tabular}
\end{indented}
\end{table}

We emphasize that such high accuracy measurements of frequency ratios are valuable inputs for the long term evaluation of optical frequency standards in view of the redefinition of the SI second \cite{Riehle2015}, as well as for long term monitoring of a putative variation of fundamental constants \cite{Huntemann2014}.

%% file: acknowledgment.tex
We acknowledge the large number of contributions of SYRTE technical services.
This work was supported by EMRP/JRP ITOC and by ERC Consolidator Grant AdOC.
We acknowledge funding support from Centre National d'\'Etudes Spatiales (CNES), Conseil R\'egional \^Ile-de-France (DIM Nano'K). The EMRP is jointly funded by the EMRP participating countries within EURAMET and the European Union.